%% file: IEEE_CM.tex
\documentclass[journal,10pt]{IEEEtran}
\usepackage{cite}
\ifCLASSINFOpdf
\else
  \usepackage[dvips]{graphicx}
  \graphicspath{{../eps/}}
  \DeclareGraphicsExtensions{.eps}
\fi
\usepackage{amsmath}
\usepackage{bm}
\usepackage{amssymb}
\usepackage{array}
\usepackage{xcolor}
\usepackage{multirow}
\usepackage{url}
\usepackage{lipsum}
\usepackage{mathtools}
\usepackage{cuted}
\usepackage{float}
\usepackage{fancyhdr}
    \usepackage{stfloats}
        \usepackage{setspace,graphicx}
    \usepackage[caption=false,font=footnotesize]{subfig}
\usepackage{stfloats}
\usepackage[linesnumbered,ruled,vlined]{algorithm2e}
\renewenvironment{IEEEbiography}[1]
  {\IEEEbiographynophoto{#1}}
  {\endIEEEbiographynophoto}
\begin{document}

\title{DFT-s-OFDM with Chirping for Integrated Sensing and Communications in 6G and Beyond}


\author{Yujie~Liu,
Yong~Liang~Guan,
David~Gonz\'alez G.,
Halim~Yanikomeroglu


\thanks{This paper was accepted for publication by IEEE Communications Magazine on 29 April 2026. Copyright (c) 20xx IEEE. Personal use of this material is permitted. Permission from IEEE must be obtained for all other uses, in any current or future media, including reprinting/republishing this material for advertising or promotional purposes, creating new collective works, for resale or redistribution to servers or lists, or reuse of any copyrighted component of this work in other works. Yujie Liu and Yong Liang Guan are with the Nanyang Technological University, Singapore. David Gonz\'alez G. is with the AUMOVIO Germany GmbH. Halim Yanikomeroglu is with the Carleton University, Canada.}}

\cfoot{\scriptsize This paper has been accepted for publication by IEEE Communications Magazine on 29 April 2026. Copyright (c) 20xx IEEE. Personal use of this material is permitted. Permission from IEEE must be obtained for all other uses, in any current or future media, including reprinting/republishing this material for advertising or promotional purposes, creating new collective works, for resale or redistribution to servers or lists, or reuse of any copyrighted component of this work in other works.}


\maketitle

\IEEEpeerreviewmaketitle

\input{paper}

\ifCLASSOPTIONcaptionsoff
  \newpage
\fi

\end{document}

%% file: paper.tex
\begin{abstract}
The sixth generation (6G) of mobile communications and beyond is expected to enable advanced functionalities, such as integrated sensing and communication (ISAC), while involving diverse terminal/user equipment types from terrestrial to non-terrestrial networks. As waveforms are acknowledged as a fundamental technology driving 6G and beyond, this article presents a contribution in this technical domain. First, it provides an overview of several standardized communication waveforms, as well as chirp-based waveforms for radar sensing and Internet of Things (IoT) applications. This article then presents single-carrier chirping waveform: discrete Fourier transform spread orthogonal frequency division multiplexing (DFT-s-OFDM) with chirping. Its fundamental principles, key properties, performances, and advantages are examined from both communication and sensing perspectives. Finally, several future research directions are outlined to further explore its potential and opportunities for ISAC.
\end{abstract}


\section{Introduction}
\begin{figure}[!t]
		\centering
		\includegraphics[width=8cm]{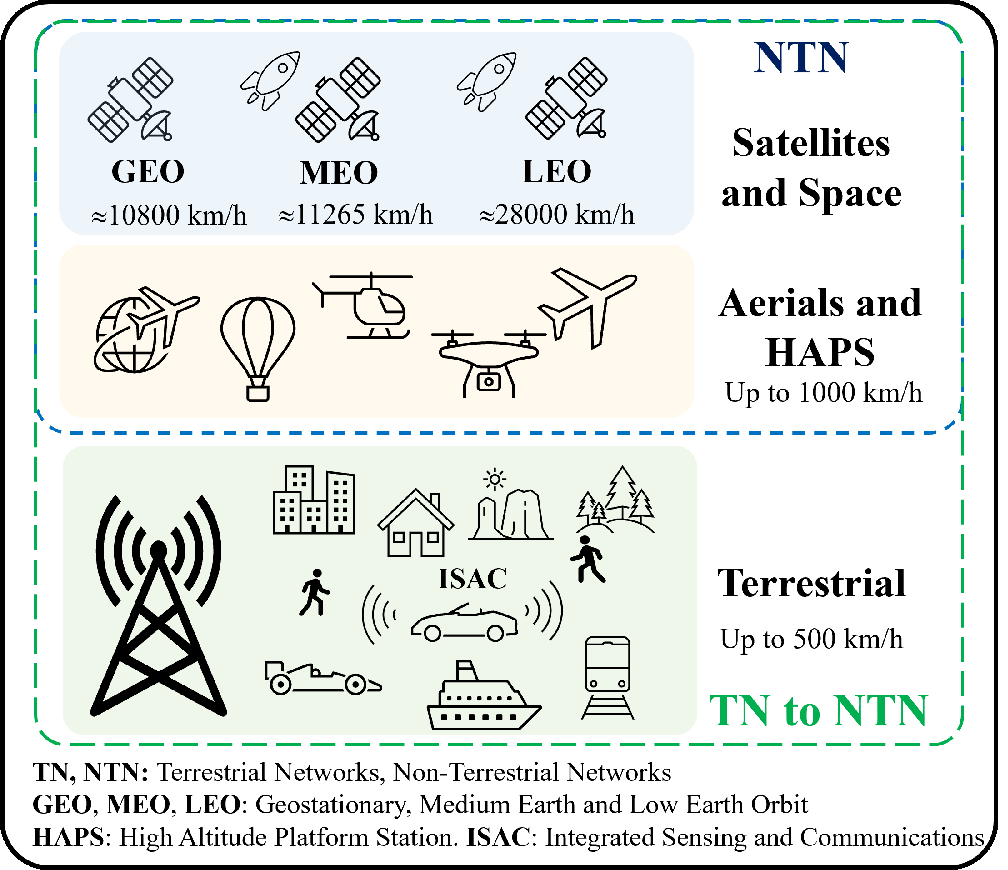}
	\caption{Toward ubiquitous connectivity, 6G and beyond communications and sensing involves diverse terminal/UE types, including power-constrained ones, TN, and NTN.}
	\label{Fig.6g_high_mobility}
\end{figure}

Over the past few decades, wireless communications has become one of the most transformative technological innovations. Since the 1980s, wireless communications evolves approximately every decade, resulting in five generations to date. The development of its sixth generation (6G) is currently in progress and its commercial release is expected to be at around 2030. As depicted in Fig. \ref{Fig.6g_high_mobility}, ubiquitous connectivity continues to be a \emph{must-have} in 6G and beyond, spanning a wide range of scenarios involving terrestrial networks \mbox{(TN)} and non-terrestrial networks (NTN). Moreover, integrated sensing and communication (ISAC) is regarded as one of key enhanced functionalities in 6G and beyond \cite{10770016,9627227,10769778}.

\begin{figure*}[!t]
		\centering
		\includegraphics[width=15cm]{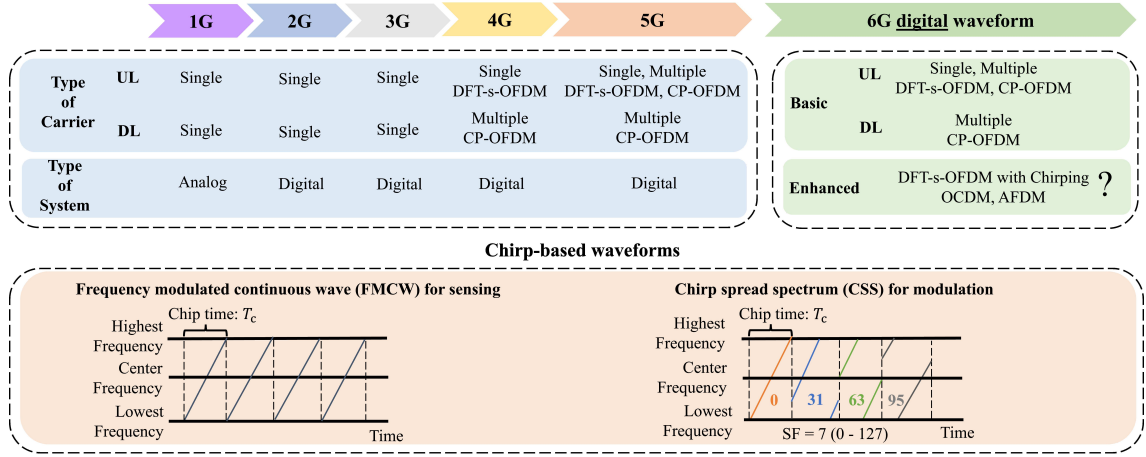}
	\caption{Overview of implemented waveforms, including the standardized waveforms from 1G to 6G and chirp-based waveforms.}
	\label{Fig.waveform}
\end{figure*}

One important feature of TN and NTN, as shown in Fig.~\ref{Fig.6g_high_mobility}, is mobility. Communication (and sensing) scenarios in TN include mobile access points and User Equipment (UE) carried by pedestrians, fast-moving vehicles, high-speed trains, etc. Mobile terminals in NTN include air taxi, uncrewed aerial vehicles (UAVs), flying high-altitude platform stations (HAPs), and swiftly orbiting satellites \cite{11112522}. The aforementioned mobility presents several challenges for 6G and beyond, including fast and frequent handovers and serious performance degradation due to Doppler effect. In addition, many mobile terminals from TN to NTN in Fig.~\ref{Fig.6g_high_mobility} are power-constrained. Hence, enhancing their power efficiency is also of paramount importance.

One of key technology component in each generation is the waveforms used for radio interfaces. The choice of waveform greatly influences the system's achievable throughput, communication reliability, spectral efficiency, power efficiency, robustness against adverse channel conditions, etc. To have high power efficiency, waveforms are supposed to have low peak-to-average-power-ratio (PAPR), where power amplifiers can be driven at its full potential (at or near their saturation points). Therefore, \emph{low} PAPR is a desirable waveform feature in 6G and beyond, and thus, this requirement has attracted significant attention from both academia and industry.

This article investigates discrete Fourier transform spread orthogonal frequency division multiplexing (DFT-s-OFDM) with chirping for ISAC in 6G and beyond. DFT-s-OFDM with chirping is formed by multiplying a DFT-s-OFDM signal with a modulated/unmodulated chirp signal. Unlike the authors' prior work \cite{10897935,liu2025dft}, which focused on proposing DFT-s-OFDM with chirping and analyzing its communication performance, this article emphasizes its ISAC capabilities, opportunities, and challenges. Section II provides an overview of standardized communication waveforms and chirp-based waveforms. Section III summarizes key communication-related insights from prior work and complements them with newly generated results. Section IV evaluates the sensing performance of DFT-s-OFDM with chirping, and Section V discusses its open research challenges and future directions for ISAC.

\section{Overview of Implemented Waveforms}
An overview of implemented waveforms utilized in communication and sensing systems is illustrated in Fig. \ref{Fig.waveform}. This section begins with an overview of standardized communication waveforms, and then reviews chirp-based waveforms.

\subsection{Standardized Waveforms from 1G to 6G}
Since the 1980s, there has been five generations of mobile communication networks, with each generation standardized every decade. The first generation (1G) was standardized in 1981 for voice communications. Analog frequency modulation is employed by modulating the voice signal onto radio frequency (RF) carriers. To improve bandwidth efficiency and data throughput, a shift from analog systems to digital communications took place in 1992. It was known as the second generation (2G). The subsequent generations have remained digital. The deployment of the third generation (3G) happened in 2000s, allowing new data and multimedia applications. The aforementioned generations of communications were all based on single carrier. To further boost network throughput, the fourth generation (4G) was the first cellular network adopting multi-carrier waveform orthogonal frequency division multiplexing (OFDM) for the downlink. However, OFDM is a superposition of multiple waveforms and their individual \emph{peaks} may be added up constructively, resulting in high PAPR. 5G also uses cyclic prefix OFDM (CP-OFDM) for downlink transmissions. Unlike 4G, CP-OFDM can also be used for 5G uplink transmissions. The single-carrier waveform based on DFT-s-OFDM can utilize spectrum more efficiently than single-carrier frequency domain equalization (SC-FDE) technique and was thus adopted as a low-PAPR waveform for 4G and 5G uplink \cite{M2516}. Since June 2025, 3GPP has made \mbox{important} agreements to consider DFT-s-OFDM and CP-OFDM as \textit{baseline} or benchmark for 6G candidate waveforms \cite{NEW}. It implies the possibility of further enhancing/modifying CP-OFDM and DFT-s-OFDM to support specific use cases or enable new capabilities \cite{NEW}.

  \begin{figure*}[!t]
	\centering
		\includegraphics[width = 15cm]{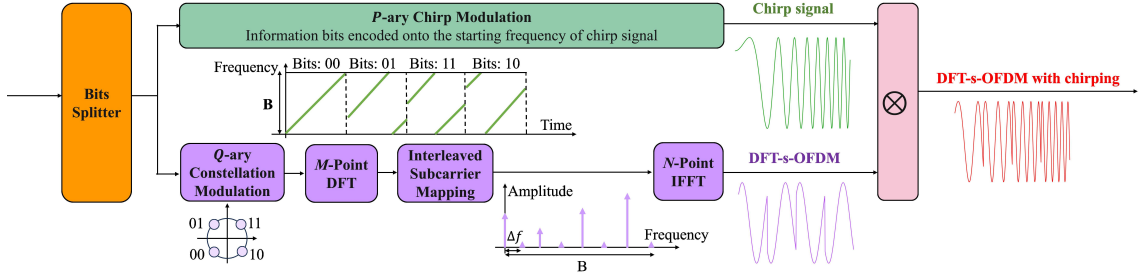}
	\caption{DFT-s-OFDM with chirping: multiplying a DFT-s-OFDM signal with either a modulated or unmodulated chirp signal.}
		\label{Fig:bd}
\end{figure*}

\subsection{Chirp-based Waveforms}
Compressed high-intensity radar pulse, also known as chirp, was invented in the 1950s, and since then has had a widespread application in numerous fields for several decades. Its important characteristics are summarized below: (a) It maintains a constant amplitude over the transmission time, resulting in low PAPR. (b) When processed with a matched filter, it yields a compressed pulse with substantially increased amplitude. (c) Due to frequency sweeping, it spans a broader frequency band and exhibits high resilience against narrow band jamming.

Thanks to the excellent properties of chirp, it has been used as the sensing waveform in radar, lidar, sonar, etc. One well-known sensing waveform is frequency modulated continuous wave (FMCW), extensively used in automotive radars, as illustrated in Fig.~\ref{Fig.waveform}. Literally speaking, FMCW emits a continuous-wave RF signal with a frequency sweep \cite{8828004}. Its principle is to mix the transmitted and received (reflected) signal, which generates a new signal with a new frequency corresponding to the frequency deviation between the transmitted and received signal. The new frequency, also known as beat frequency, is used for range and altitude measurements.

In addition to sensing, chirping can also been used for communications. LoRa (long range) is based on chirp spread spectrum (CSS) modulation, targeting for long-range communications especially with power-constrained Internet of Things (IoT) devices \cite{10323409}. An illustrative example is provided in Fig. \ref{Fig.waveform} to depict how to encode data onto chirps. LoRa can demodulate signals with power $20\,\rm{dB}$ below the noise level, making it robust in communication scenarios with severe interference. Despite of the aforementioned advantage, LoRa communications have extremely low data rate, making it unsuitable for broadband applications requiring high data rates.

ISAC is regarded as a key usage scenario for 6G and beyond. Recent research efforts have focused on integrating communication waveforms with chirp-based waveforms. Specifically, orthogonal chirp division multiplexing (OCDM)~\cite{10136610} and affine frequency division multiplexing (AFDM)~\cite{10087310} were proposed by performing frequency-domain and time-domain chirping before and after OFDM. Orthogonal time frequency space (OTFS) \cite{10769778} and its variant, orthogonal delay scale space (ODSS) \cite{9772941}, have been proposed with superior communication performance. The sensing capability of OTFS and AFDM was compared and discussed in \cite{10769778}. A cross-layer vision for ISAC in 6G was presented in \cite{11112522}, while \cite{10770016} reviewed multicarrier waveforms and signal processing techniques for ISAC, and \cite{9627227} compared several state-of-the-art modulation schemes in ISAC scenarios. These works provide valuable insights on waveform design for ISAC, but the considered waveforms generally suffer from high PAPR. In contrast, this paper focuses on the single-carrier chirping waveform DFT-s-OFDM with chirping and evaluates its ISAC capabilities, opportunities, and challenges, primarily at the physical layer.

\section{DFT-s-OFDM with Chirping: Communications}
This section presents the fundamental principles and key properties of DFT-s-OFDM with chirping, from a communications perspective. Fig. \ref{Fig:bd} illustrates the block diagram of DFT-s-OFDM with chirping. The main concept involves multiplying the DFT-s-OFDM signal with either a modulated or an unmodulated chirp signal. DFT-s-OFDM consists of an $M$-point discrete Fourier transform (DFT), a subcarrier mapping scheme, and an $N$-point inverse fast Fourier transform (IFFT). Interleaved subcarrier mapping is considered throughout this article. Since $M$ is typically smaller than $N$, DFT-s-OFDM uses a subset of interleaved subcarriers, as illustrated in the frequency-domain DFT-s-OFDM signal. The DFT-s-OFDM symbol remains the original constellation symbol, with reduced amplitude and phase rotation \cite{9143507}. Hence, its analog waveform exhibits single-carrier characteristic in Fig. \ref{Fig:bd}.

A chirp waveform is commonly depicted as a line representing a linear frequency sweep over time across the bandwidth $B$, where the starting point corresponds to the starting frequency. When an unmodulated chirp signal is used, the resulting waveform is named as chirped DFT-s-OFDM \cite{10897935}. If a modulated chirp signal is employed, where information bits are conveyed by modulating its starting frequency, the resulting waveform is termed DFT-s-OFDM-CM \cite{liu2025dft}. In the middle of Fig. \ref{Fig:bd}, four chirp waveforms are depicted as four groups of green lines, with each group having a distinct starting point that corresponds to a different information bit sequence, \emph{e.g.}, $00$, $01$, $11$, and $10$. See Section II in \cite{liu2025dft} for more technical details. Fig. \ref{Fig:bd} illustrates an example of the analog chirp waveform corresponding to the bit sequence $00$, where the chirp starts from the lowest frequency. After multiplying the DFT-s-OFDM signal with chirp signal, the resulting DFT-s-OFDM with chirping preserves the single-carrier characteristic of DFT-s-OFDM, while exhibiting the frequency-sweep nature of chirp. Its key properties are summarized as follows.

  \begin{figure*}[htbp]
	\centering
	\includegraphics[width = 18cm]{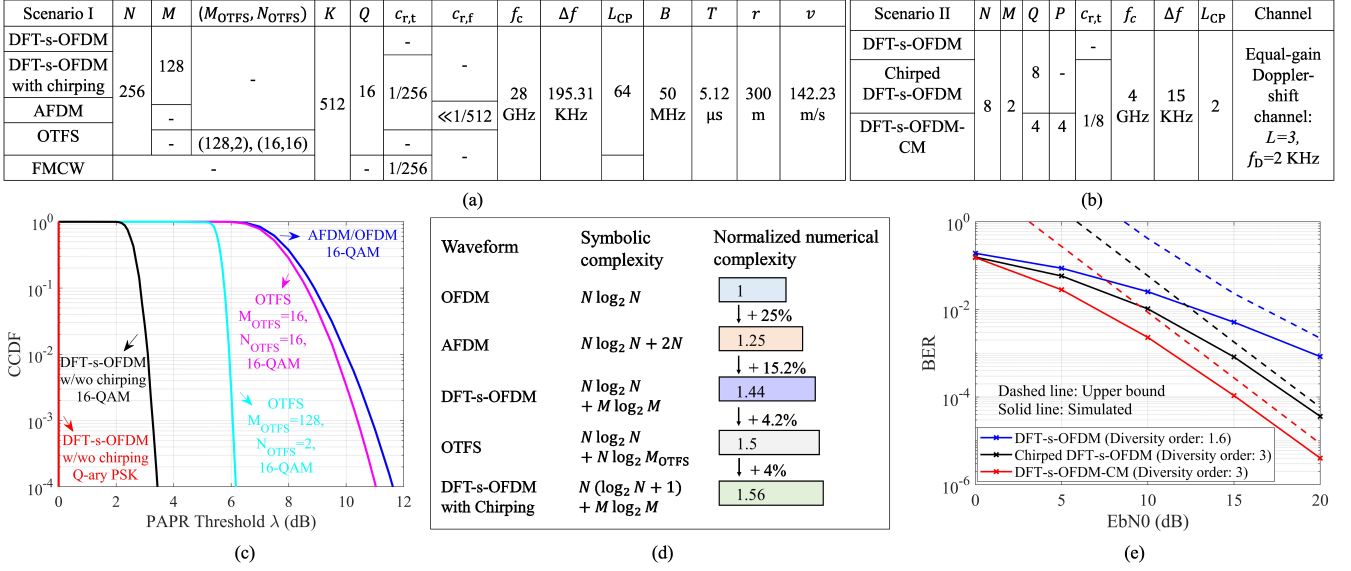}
	\caption{(a) Simulation setup: scenario I \cite{10087310,10102330}. (b) Simulation setup: scenario II \cite{10897935,liu2025dft}. (c) PAPR. (d). Modulation complexity. (e). BER.}
	\label{Fig:comm}
\end{figure*}

 \textbf{Spectral efficiency}: In this paper, DFT-s-OFDM with and without chirping are compared to commonly used AFDM, OFDM, and OTFS systems which employ all subcarriers for transmission. Since DFT-s-OFDM and its chirped variant transmit constellation symbols over a subset $(M/N)$ of subcarriers, they usually present lower spectral efficiency than AFDM, OFDM, and OTFS. Nevertheless, under this configuration, chirped DFT-s-OFDM has been shown in \cite{10897935} to achieve better noise suppression. The spectral efficiencies of DFT-s-OFDM and chirped DFT-s-OFDM are the same and expressed as $\frac{M\log_2{Q}}{N}$, where $M$, $N$, and $Q$ are respectively DFT size, IFFT size, and constellation modulation order. After incorporating chirp modulation, the spectral efficiency of DFT-s-OFDM-CM can increase to $\frac{M\log_2{Q}+\log_2{P}}{N}$, where $P$ is chirp modulation order. Additionally, subcarrier index modulation, in which the active subcarrier indices in DFT-s-OFDM carry additional bits, can be used to further improve spectral efficiency. As in multi-access systems, the remaining $(N-M)$ subcarriers in DFT-s-OFDM can be allocated to other users for data transmission.

\textbf{Low PAPR}: This paper investigates symbol-level PAPR, calculated in discrete time with an oversampling ratio of $1$ and rectangular pulse shaping. Chirp primarily changes frequency, without affecting the amplitude and PAPR. For DFT-s-OFDM with chirping, its PAPR is thus determined by DFT-s-OFDM parameters, including subcarrier mapping and constellation modulation. Interleaved mapping is considered throughout this paper, and the resulting PAPR is independent of DFT size $M$ and FFT size $N$, corresponding to that of the original constellation symbols \cite{9143507,10897935}. In contrast, the PAPR of OFDM/AFDM scales with the number of subcarriers, whereas OTFS is influenced by the number of Doppler grids. Following the simulation setups in \cite{10087310} and \cite{10102330} as shown in Fig.~\ref{Fig:comm}a, the complementary cumulative distribution function (CCDF) of PAPR for DFT-s-OFDM w/wo chirping are plotted in Fig.~\ref{Fig:comm}c against OFDM, AFDM, and OTFS. The definition of CCDF of PAPR is the probability that PAPR is above a certain value, \emph{i.e.}, $\rm{CCDF}=\rm{Pr}(\rm{PAPR}>\lambda)$. DFT-s-OFDM w/wo chirping has a PAPR of approximately $\lambda=3.5\,\rm{dB}$ to achieve CCDF of $10^{-4}$, while $\lambda>6\,\rm{dB}$ is required for OFDM, AFDM, and OTFS. Moreover, the PAPR of DFT-s-OFDM w/wo chirping decreases to $0\,\rm{dB}$ if phase shift keying (PSK) modulation is adopted, regardless of the modulation order. The PAPR of OTFS decreases as $N_{\textrm{OTFS}}$ reduces from $16$ to $2$. Unless otherwise stated, OTFS configuration with $M_{\textrm{OTFS}}=128$ and with $N_{\textrm{OTFS}}=2$ is used throughout the paper thanks to its lower PAPR. Note that under a multi-access scheme, OFDM, AFDM, and OTFS using interleaved subcarriers or delay grids still show higher PAPR than DFT-s-OFDM. The corresponding results are omitted due to space limitations.

\textbf{Modulation Complexity}: Fig. \ref{Fig:comm}d compares the transmitter-side modulation complexity of DFT-s-OFDM w/wo chirping, AFDM, OTFS, and OFDM. The more common implementation of OTFS modulation using FFT and IFFT \cite{10087310,10769778} is considered. Using the setup in Fig.~\ref{Fig:comm}a, the complexities of all waveforms are calculated and normalized to that of OFDM. Due to extra processing before and after the IFFT, DFT-s-OFDM with chirping incurs higher complexity than OFDM and DFT-s-OFDM. The complexity difference between OTFS and chirped DFT-s-OFDM arises from the Heisenberg transform in OTFS versus the $M$-point DFT and chirping operations in chirped DFT-s-OFDM, and can be expressed as $N\log_2M_{\rm{OTFS}}-M\log_2M-N$. This difference depends on $M_{\rm{OTFS}}$, $M$, and $N$. For example, using the setup in Fig. \ref{Fig:comm}a with $M_{\rm{OTFS}}=16$ and $N_{\rm{OTFS}}=16$, DFT-s-OFDM with chirping exhibits approximately a $4\%$ increase in transmitter-side complexity over OTFS. Nevertheless, there are also some scenarios in which OTFS exhibits higher modulation complexity, for example, for fixed $N$ and $M$, selecting $M_{\rm{OTFS}}\geq 32$. The normalized modulation complexity of OTFS with $M_{\rm{OTFS}}=128$ and $N_{\rm{OTFS}}=2$ is calculated as $1.88$, which is around $21\%$ higher than that of DFT-s-OFDM with chirping. Similarly, the complexity difference between AFDM and DFT-s-OFDM with chirping is written as $N-M\log_2M$. AFDM may also exhibit higher complexity than chirped DFT-s-OFDM, for example when $N=256$ and $M\leq32$. Moreover, as the DFT-s-OFDM symbol remains as the original constellation symbol with reduced amplitude and phase rotation, the DFT, subcarrier mapping, and IFFT operations can be replaced by repetition and phase-shift modulation to reduce complexity \cite{9143507}. Receiver demodulation follows a similar complexity trend and is omitted; see \cite{10897935} for details.

\textbf{Receiver Implementation:}
There are two cases for receiver-side processing of DFT-s-OFDM with chirping in communication systems. When no chirp modulation is applied, the receiver design for chirped DFT-s-OFDM is relatively straightforward. Similar to AFDM, the chirp signal is assumed to be known at the receiver and is incorporated into the channel matrix to form an equivalent channel representation. Then, the DFT-s-OFDM signal can be recovered through standard equalization techniques, such as maximum-likelihood (ML), linear minimum mean square error (LMMSE), message passing (MP), etc. In this case, DFT-s-OFDM with chirping may exhibit similar equalization complexity to other waveforms. When chirp modulation is applied, the DFT-s-OFDM and chirp signals are mixed and mutually interfere with each other. As a preliminary research, ML receiver designs developed for OFDM/AFDM with index modulation can be adopted, in which the chirp signal and DFT-s-OFDM signal are jointly estimated via an exhaustive search over all possible combinations. Its receiver complexity is comparable to OFDM/AFDM with index modulation. While ML detection is optimal, its relatively higher complexity motivates the development of low-complexity receivers for DFT-s-OFDM-CM.

  \begin{figure*}[htbp]
	\centering
		\includegraphics[width = 18cm]{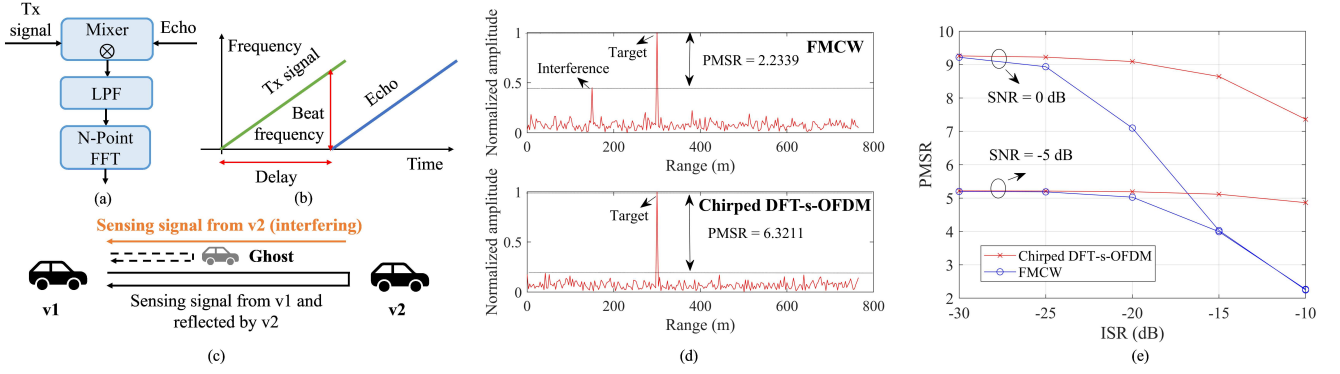}
	\caption{(a) Analog mixer. (b) Illustrative time-frequency diagram of transmitted signal and its echo. (c) Single-target sensing in the presence of interference. (d) Range ambiguity function. (e) PMSR as a function of ISR.}
		\label{Fig:sensing_st}
\end{figure*}

\textbf{Full Frequency Diversity}: Since ML detection suffers high computational complexity, it is typically applied to small-scale systems \cite{10897935}. Using the simulation setup in Fig. \ref{Fig:comm}b \cite{10897935,liu2025dft}, the BERs of DFT-s-OFDM and its chirped variants are obtained in Fig. \ref{Fig:comm}e using ML equalizer. Fig. \ref{Fig:comm}e also shows their BER upper bounds, which are determined using their pairwise error probability (PEP) analysis of ML in \cite{10897935} and \cite{liu2025dft}.
It demonstrates that DFT-s-OFDM with chirping, including chirped DFT-s-OFDM and DFT-s-OFDM-CM, achieves full frequency diversity corresponding to the number of propagation paths ($L=3$), similar to AFDM and OTFS. In contrast, DFT-s-OFDM without chirping fails to fully exploit frequency diversity. In addition, the superior BER performance of DFT-s-OFDM-CM over chirped DFT-s-OFDM stems from splitting the data bits into two parts. Under the same spectral efficiency, this enables DFT-s-OFDM-CM to use a lower-order constellation, thereby improving robustness to noise \cite{liu2025dft}.

\section{DFT-s-OFDM with Chirping: Sensing}
This section evaluates the sensing capabilities and performances of DFT-s-OFDM with chirping using two common radar sensing techniques: analog-domain mixing and digital-domain matched filtering, in terms of ambiguity functions, peak-to-max-side lobe ratio (PMSR), probability of detection, range and Doppler resolution. Note that chirped DFT-s-OFDM and DFT-s-OFDM-CM have identical sensing performance, as the chirping modulation affects only communication. Accordingly, they are not distinguished in this section, and all related results are based on chirped DFT-s-OFDM.


\subsection{Analog Mixing}

According to Fig. \ref{Fig:bd}, DFT-s-OFDM with chirping preserves frequency sweeping nature of chirp signal. Additionally, by applying a DFT prior to OFDM, together with interleaved subcarrier mapping, DFT-s-OFDM becomes a single-carrier waveform \cite{9143507}. Hence, similar to FMCW, DFT-s-OFDM with chirping can be processed using an analog mixer to generate a beat frequency for target range estimation. Fig. \ref{Fig:sensing_st}a illustrates the block diagram of analog mixer. The transmitted chirped DFT-s-OFDM signal and its received echo are mixed as its two inputs. The output of the analog mixer contains two components: one at the frequency difference (beat frequency) and another at the frequency sum. The low pass filter (LPF) is required to suppress the sum-frequency component, after which an $N$-point FFT is performed. The locations of the peaks in the FFT magnitude are then used to determine the beat frequency \cite{8828004}. Subsequently, the target range can be calculated using the beat frequency and chirp rate according to the time-frequency diagram in Fig. \ref{Fig:sensing_st}b. Note that although the original waveform is not explicitly present after mixing, the sensing information is preserved in the mixed signal. Unlike FMCW, which carries no data and supports sensing only, the chirped DFT-s-OFDM signal enables simultaneous sensing and communications. Additionally, its unique and independent data sequence helps mitigate sensing interference.

Fig. \ref{Fig:sensing_st}c depicts a sensing scenario to detect a target when interference is present. As shown in Fig. \ref{Fig:sensing_st}c, both Vehicle $1$ and Vehicle $2$ are transmitting its sensing signal for target detection. In addition to receiving its own echo reflected by Vehicle $2$, Vehicle $1$ also receives the sensing signal transmitted by Vehicle 2, shown as interference in the orange line. Following the simulation setup in Fig. \ref{Fig:comm}a with the signal-to-noise ratio (SNR) set to $-5\,\rm{dB}$ and interference-to-signal ratio (ISR) to $-10\,\rm{dB}$, the range ambiguity functions using FMCW and chirped DFT-s-OFDM are shown in Fig. \ref{Fig:sensing_st}d. Since the FMCW signal carries no data and both vehicles transmit the same waveform, the signal from Vehicle 2 is mistakenly interpreted as an echo from a non-existent target (a ghost target). This produces two peaks in the range ambiguity function, one of which corresponds to the ghost target. Rather than using the same FMCW signal, the two vehicles each transmit distinct chirped DFT-s-OFDM signals, with each signal carrying different information bits. Since the data transmitted by each vehicle are likely to differ, the two chirped DFT-s-OFDM signals are likely to be uncorrelated. When the received superimposed signal is mixed with Vehicle $1$'s transmitted signal, the signal transmitted from Vehicle $2$ could be suppressed due to its low correlation.
This is also why chirped DFT-s-OFDM presents higher PMSR than FMCW in Fig. \ref{Fig:sensing_st}d. Their PMSRs are also simulated as a function of ISR in Fig. \ref{Fig:sensing_st}e using the setup in Fig. \ref{Fig:comm}a. When ISR is low, both waveforms exhibit similar PMSRs. As the ISR increases, however, the PMSR of FMCW degrades more rapidly than that of chirped DFT-s-OFDM. This indicates that DFT-s-OFDM with chirping is more robust to interference than FMCW.


  \begin{figure*}[!t]
	\centering
		\includegraphics[width = 18cm]{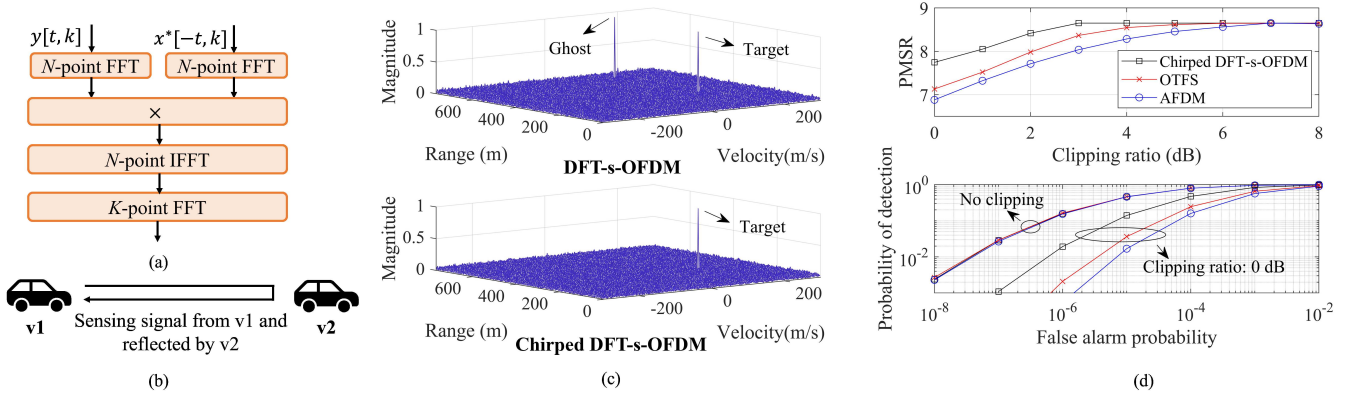}
	\caption{(a) Matched filter. (b) Single-target sensing. (c) Range-velocity ambiguity functions. (d) PMSR and probability of detection w/wo clipping.}
	\label{Fig.fmcw_chirped}
\end{figure*}

\subsection{Matched Filtering}

Multicarrier chirping waveforms including AFDM and OCDM cannot be effectively processed by analog mixing. Since multiple carriers produce multiple beat frequencies, it is difficult to identify the beat frequency corresponding to the target range. Moreover, waveforms that do not incorporate chirping, including OFDM, OTFS, and DFT-s-OFDM, are also unsuitable for analog mixing. Since their carrier frequencies do not sweep over time, the beat-frequency property required for range estimation does not exist. To facilitate a comparison among the aforementioned waveforms in terms of sensing performance, matched filter technique is employed.

The block diagram of matched filter is shown in Fig. \ref{Fig.fmcw_chirped}a. A matched filter correlates the received signal with the time-reversed conjugate of the transmitted waveform, expressed as $y(n,k)\ast x^{\ast}(-n,k)$, where $n$ and $k$ respectively correspond to fast-time and slow-time sample index \cite{9627227,8828004}. Note that $y(n,k)\ast x^{\ast}(-n,k)$ can be obtained equivalently using two $N$-point FFTs, one multiplication, and one $N$-point IFFT, with $N$ being the number of fast-time samples or total subcarriers. A subsequent $K$-point FFT in the slow-time domain is then applied to obtain the joint range-velocity ambiguity function, with $K$ being the number of slow-time samples \cite{9627227}.

Fig. \ref{Fig.fmcw_chirped}b illustrates a single sensing target scenario, where Vehicle $1$ transmits a signal (e.g., DFT-s-OFDM, chirped DFT-s-OFDM, AFDM, OTFS) and receives the echo reflected by Vehicle~$2$. Based on the simulation setup in Fig. \ref{Fig:comm}a with SNR of $-20\,\rm{dB}$, the range-velocity ambiguity functions of DFT-s-OFDM and its chirped variant are obtained in Fig. \ref{Fig.fmcw_chirped}c. Due to DFT precoding/spreading with $N/M=2$, DFT-s-OFDM symbols are repeated twice with reduced amplitude, resulting in two peaks. This repetition is avoided when chirping is applied, and thus, only a single peak appears for chirped DFT-s-OFDM. In practice, high-power input signals cause the power amplifier to hit its maximum output, resulting in clipping \cite{10897935}. Considering the setup in Fig. \ref{Fig:comm}a, the PMSRs and probability of detection of various waveforms are simulated with and without clipping in Fig. \ref{Fig.fmcw_chirped}d, with SNR set to $-20\,\rm{dB}$ and $-40\,\rm{dB}$, respectively. Without clipping, chirped DFT-s-OFDM achieves PMSR and detection probability comparable to AFDM and OTFS. As shown in Fig. \ref{Fig:comm}c, AFDM and OTFS exhibit higher PAPR and are therefore more sensitive to clipping than chirped DFT-s-OFDM. Their PMSR and detection probability are found to be worse than chirped DFT-s-OFDM, as clipping ratio decreases.

With analog mixing and matched filtering, range and velocity resolution depend on the transmission parameters and are independent of the waveform. According to \cite{9627227}, the range resolution is determined by the bandwidth $B$ and given by $\frac{c}{2B}$, where $c$ is the speed of light. The velocity resolution then depends on carrier frequency $f_{\textrm{c}}$, bandwidth $B$, fast-time length $(N+L_{\textrm{CP}})$, and slow-time length $K$ , and is given by $\frac{Bc}{2f_{\textrm{c}}(N+L_{\textrm{CP}})K}$. Under the simulation setup in Fig. \ref{Fig:comm}a, the range and velocity resolution are calculated as $3\,\rm{m}$ and $1.63\,\rm{m/s}$.

\section{Open Challenges and Future Directions}
Despite its advantages for ISAC, DFT-s-OFDM with chirping still faces several challenges and open issues.

\textbf{Chirp Shape Analysis and Optimization:}
Chirping is one important feature of DFT-s-OFDM with chirping. Linear chirp with frequency increasing/decreasing linearly over time is usually considered for simplicity. In addition to linear chirps, various nonlinear chirps, such as hybrid, quadratic, cubic, exponential, logarithmic, and sinusoidal shapes, have also been proposed. The impact of chirp shape should be analyzed and optimized to trade-off communication reliability, sensing resolution, spectral efficiency, latency, interference, etc.

\textbf{Chirp Modulation and Index Modulation:}
As described earlier, DFT-s-OFDM can be integrated with chirp modulation and index modulation to enhance spectral efficiency. Preliminary studies~\cite{liu2025dft} have demonstrated the effectiveness of DFT-s-OFDM with chirp modulation. However, the combined effects of chirp and index modulation on DFT-s-OFDM remain unclear. Their joint use may introduce ambiguity issues. 

\textbf{Multiuser Interference Management:}
In multi-access communications, DFT-s-OFDM maintains orthogonality among multiple users. However, when chirping is applied, each user's frequency sweeps into other users' bands, resulting in multiuser interference. ML, as an optimal equalizer, is able to eliminate multiuser interference. However, its high complexity limits its practical implementation. Advanced interference management techniques need to be developed.

\textbf{Non-Terrestrial Networks:}
NTN represents a major research focus of 6G and beyond. The NTN standardized framework encompasses satellites, HAPS, and UAVs. These aerial platforms are typically power-limited and the high relative velocity between satellites and users causes significant Doppler effects. Introducing chirping into DFT-s-OFDM maintains low PAPR and enhances its robustness against Doppler shifts, making it a promising waveform for NTN.

\textbf{AI Empowered Research:}
Artificial intelligence (AI)/machine learning (ML) approaches are becoming deeply involved in 6G and beyond standardization for waveform optimization, hardware impairment compensation, channel estimation, and signal detection. Introducing AI/ML methods to DFT-s-OFDM with chirping can enhance its performance, reliability, adaptability, and efficiency, simplify receiver design, and reduce implementation complexity. 

\textbf{Practical Receiver Issues:}
Chirping can complicate receiver design. Frequency sweeps may couple timing and frequency offsets, which is further worsen by Doppler in high-mobility and NTN scenarios. Practical receivers may require dechirping before synchronization or joint estimation of timing offset, frequency offset, Doppler shift, and chirp rate. Additionally, chirping poses challenges for channel estimation, as the approaches developed for DFT-s-OFDM do not directly extend to chirped waveforms. Similar to AFDM~\cite{10087310}, element-wise or basis-expansion models that incorporate Doppler effects need to be derived to enable pilot design and channel estimation for DFT-s-OFDM with chirping. Digital implementation of chirping as in AFDM and OCDM is generally preferred for simplicity, and it does not impact phase noise. However, when chirping is applied in the analog domain as in FMCW, frequency sweeping in a phase-locked loop can introduce higher phase noise than operation at a fixed carrier. Its impact and compensation require further investigation.

\textbf{Advanced Digital Sensing Techniques:}
The root mean square error (RMSE) of range and velocity estimation is not commonly used in analog mixing and matched filtering, but it also remains an important sensing performance metric. RMSE-based metrics are typically associated with ML estimators, high-resolution algorithms, such as Multiple Signal Classification (MUSIC) and Estimation of Signal Parameters via Rotational Invariance Techniques (ESPRIT), or Cr\'amer-Rao Lower Bound (CRLB) analysis. These advanced digital sensing techniques and analyses provide valuable insights and are worth further exploration in future work.

The discussion above focuses on DFT-s-OFDM with chirping for ISAC. Other emerging waveforms, including OCDM~\cite{10136610}, AFDM~\cite{10087310}, ODSS~\cite{9772941}, and OTFS~\cite{10769778}, are also promising for ISAC and worth exploring in future research.

\section{Conclusion}
This article presents an overview of commercially available waveforms, including standardized communication waveforms as well as chirp-based waveforms for radar sensing and IoT applications. The fundamental principles and properties of DFT-s-OFDM with chirping are then discussed. DFT-s-OFDM with chirping could retain the low PAPR characteristic of DFT-s-OFDM while achieving full frequency diversity similar to AFDM and OTFS. The waveform also offers high flexibility for integration with chirp and/or index modulation to enhance spectral efficiency. As a single-carrier chirping waveform, it is compatible with existing radar receivers using analog mixing and can also suppress ghost targets caused by interference. It is also more robust to clipping than AFDM and OTFS and eliminates false peaks from DFT precoding.

\section*{Acknowledgment}
This research/project is supported by A*STAR under the RIE2025 Industry Alignment Fund - Industry Collaboration Projects (IAF-ICP) Funding Initiative (Award: I2501E0045), as well as cash and in-kind contribution from the industry partner(s).

\bibliographystyle{IEEEtran}
{\small
\bibliography{references}}
\addcontentsline{toc}{section}{References}

\vskip -2.5\baselineskip plus -1fil

\begin{IEEEbiography}{Yujie Liu} [M] (yujie.liu@ieee.org) is a Senior Research Fellow at the AUMOVIO-NTU (formerly Continental-NTU) Corporate Research Lab, Nanyang Technological University, Singapore. 
\end{IEEEbiography}
\vskip -2.7\baselineskip plus -1fil
\begin{IEEEbiography}{Yong Liang Guan} [SM] (eylguan@ntu.edu.sg) is an Associate Vice President and a Professor at Nanyang Technological University, Singapore. He leads the AUMOVIO-NTU (formerly Continental-NTU) Corporate Research Lab and led the successful deployment of the campus-wide NTU-NXP V2X Test Bed.
\end{IEEEbiography}
\vskip -2.7\baselineskip plus -1fil
\begin{IEEEbiography}{David Gonz\'alez G.} [SM] (david.gonzalez.g@ieee.org) is with AUMOVIO Germany GmbH. He leads research projects on 5G-A and 6G, and participates as a delegate in 3GPP RAN1, 5GAA, and ETSI ISG ISAC.
\end{IEEEbiography}
\vskip -2.7\baselineskip plus -1fil
\begin{IEEEbiography}{Halim~Yanikomeroglu} [F] (halim@sce.carleton.ca) is a Chancellor's Professor at Carleton University, Ottawa, Ontario, Canada. He is with the Department of Systems and Computer Engineering and is the Founding Director of the Carleton-NTN (Non-Terrestrial Networks) Lab.
\end{IEEEbiography}


%% file: references.bib
@ARTICLE{9143507,
  author={Shao, Yulin and Liew, Soung Chang},
  journal={IEEE Trans. Wireless Commun.},
  title={Flexible Subcarrier Allocation for Interleaved Frequency Division Multiple Access},
  year={2020},
  volume={19},
  number={11},
  pages={7139-7152},
  doi={10.1109/TWC.2020.3008484}}

@ARTICLE{10770016,
  author={Koivunen, Visa and Keskin, Musa Furkan and Wymeersch, Henk and Valkama, Mikko and Gonz\'alez-Prelcic, Nuria},
  journal={IEEE Signal Process. Mag.},
  title={Multicarrier {ISAC}: Advances in waveform design, signal processing, and learning under nonidealities},
  year={2024},
  volume={41},
  number={5},
  pages={17-30},
  doi={10.1109/MSP.2024.3420492}}

@ARTICLE{11112522,
  author={Wymeersch, Henk and et al.},
  journal={IEEE Open J. Commun. Soc.},
  title={Cross-Layer Integrated Sensing and Communication: A Joint Industrial and Academic Perspective},
  year={2025},
  volume={6},
  number={},
  pages={6966-7015},
  doi={10.1109/OJCOMS.2025.3595459}}

@ARTICLE{10102330,
  author={Keskin, Musa Furkan and Wymeersch, Henk and Koivunen, Visa},
  journal={IEEE Trans. Signal Process.},
  title={Monostatic Sensing With {OFDM} Under Phase Noise: From Mitigation to Exploitation},
  year={2023},
  volume={71},
  number={},
  pages={1363-1378},
  doi={10.1109/TSP.2023.3266976}}

@ARTICLE{10769778,
  author={Rou, Hyeon Seok and et al.},
  journal={IEEE Sig. Process. Mag.},
  title={From Orthogonal Time-Frequency Space to Affine Frequency-Division Multiplexing: A comparative study of next-generation waveforms for integrated sensing and communications in doubly dispersive channels},
  year={2024},
  volume={41},
  number={5},
  pages={71-86},
  doi={10.1109/MSP.2024.3422653}}

@ARTICLE{9772941,
  author={K. P., Arunkumar and Murthy, Chandra R.},
  journal={IEEE Trans. Signal Process.},
  title={Orthogonal Delay Scale Space Modulation: A New Technique for Wideband Time-Varying Channels},
  year={2022},
  volume={70},
  number={},
  pages={2625-2638},
  doi={10.1109/TSP.2022.3174415}}

@misc{NEW,
         TITLE = "{ R1-2508043: Feature Lead summary \#3 on 6G waveform (Source: Nokia)}",
         AUTHOR= "{3GPP TSG RAN}",
         MONTH= oct,
         YEAR= "2025",
         NOTE= "{RAN1 Meeting \#122bis}"}

@ARTICLE{10136610,
  author={Ouyang, Xing and Dobre, Octavia A. and Guan, Yong Liang and Townsend, Paul},
  journal={IEEE Trans. Wireless Commun.},
  title={Channel Estimation for Multiple-Input Multiple-Output Orthogonal Chirp-Division Multiplexing Systems},
  year={2024},
  volume={23},
  number={1},
  pages={436-449},
  doi={10.1109/TWC.2023.3278173}}

@ARTICLE{10897935,
  author={Liu, Yujie and Guan, Yong Liang and Gonz\'alez G., David and Yanikomeroglu, Halim},
  journal={IEEE Tran. Veh. Technol.},
  title={{Chirped DFT-s-OFDM}: A New Single-Carrier Waveform With Enhanced {LMMSE} Noise Suppression},
  year={2025},
  volume={74},
  number={7},
  pages={11501-11506},
  doi={10.1109/TVT.2025.3544337}}

@INPROCEEDINGS{liu2025dft,
  author={Liu, Yujie and Guan, Yong Liang and Gonz\'alez G., David  and Yanikomeroglu, Halim},
  booktitle={Proc. IEEE 36th Int. Symp. Pers., Indoor Mobile Radio Commun. (PIMRC)},
  title={{DFT-s-OFDM} with Chirp Modulation},
  year={2025},
  volume={},
  number={},
  pages={1-6}}

@ARTICLE{10087310,
	author={Bemani, Ali and Ksairi, Nassar and Kountouris, Marios},
	journal={IEEE Trans. Wireless Commun.},
	title={Affine Frequency Division Multiplexing for Next Generation Wireless Communications},
	year={2023},
	volume={22},
	number={11},
	pages={8214-8229},
	doi={10.1109/TWC.2023.3260906}}

@ARTICLE{8828004,
  author={Hakobyan, Gor and Yang, Bin},
  journal={IEEE Signal Process. Mag.},
  title={High-Performance Automotive Radar: A Review of Signal Processing Algorithms and Modulation Schemes},
  year={2019},
  volume={36},
  number={5},
  pages={32-44},
  doi={10.1109/MSP.2019.2911722}}

@ARTICLE{9627227,

	author={Giroto de Oliveira, Lucas and Nuss, Benjamin and Alabd, Mohamad Basim and Diewald, Axel and Pauli, Mario and Zwick, Thomas},

	journal={IEEE Tran. Micro. Theory Techn.},

	title={Joint Radar-Communication Systems: Modulation Schemes and System Design},

	year={2022},

	volume={70},

	number={3},

	pages={1521-1551},

	doi={10.1109/TMTT.2021.3126887}}

@ARTICLE{10323409,
	
	author={Kietzmann, Peter and Alamos, José and Kutscher, Dirk and Schmidt, Thomas C. and Wählisch, Matthias},
	
	journal={IEEE Commun. Mag.},
	
	title={Rethinking {LoRa} for the {IoT}: An InformationCentric Approach},
	
	year={2024},
	
	volume={62},
	
	number={1},
	
	pages={34-39},
	
	doi={10.1109/MCOM.001.2300379}}

@misc{M2516,
	TITLE = "{Report M.2516-0: Future technology trends of terrestrial International Mobile Telecommunications systems towards 2030 and beyond}",
	AUTHOR = "{ITU-R}",
	MONTH = {Nov.},
	YEAR = {2022},
NOTE="{Available online~at:~\url{https://www.itu.int/dms_pub/itu-r/opb/rep/R-REP-M.2516-2022-PDF-E.pdf}}"}
